\RequirePackage{fix-cm}
\documentclass[twocolumn,epjc3]{svjour3}  
\smartqed  % flush right qed marks, e.g. at end of proof
\RequirePackage{graphicx}
\usepackage{amsmath}
\usepackage{mathrsfs}

\usepackage{fix-cm} 
\usepackage{xcolor}

\journalname{Eur. Phys. J. C}
\begin{document}

\title{Thermodynamic Geodesics in Bardeen Regular Black Hole: Conventional vs. Modified Geometrothermodynamics Metrics}
% \subtitle{Do you have a subtitle?\\ If so, write it here}


\author{Gunindra Krishna Mahanta\thanksref{e1,addr1,addr2}}

\thankstext{e1}{e-mail: guninmohantaba@gmail.com}


\institute{Astrophysical Sciences Division, Bhabha Atomic Research Centre, Mumbai, 400085, Maharashtra, India \label{addr1}
           \and
           Homi Bhabha National Institute, Anushaktinagar, Mumbai, 400094, Maharashtra, India \label{addr2}
           % \and
           % \emph{Present Address:} if needed\label{addr3}
}

\date{Received: date / Accepted: date}

\maketitle

\begin{abstract}
Thermodynamic geometry allow us to study the microscopic behavior of black hole system  by defining a metric structure in thermodynamic phase space.  Among the various thermodynamic metric structures, metrics defined by geometrothermodynamics (GTD) are extensively used to study the various thermodynamic system due to its Legendre invariant nature. In this work we investigate the behavior of thermodynamic geodesic of Bardeen regular black hole in thermodynamic space defined by three
different GTD metrics. Based on the behavior of thermodynamic geodesic as well as thermodynamic
curvature we argued that conventional GTD metric need some modifications to reflect all the thermodynamical properties of a system. We also modified the conventional GTD metrics and explore
the behavior of thermodynamic geodesic defined by the modified metrics. Our study shows that the modified GTD metrics contain most of the information about the thermodynamical boundaries such as temperature vanishing line, spinodal line etc. of a black hole system. Based on the property of geodesic and Ricci scalar defined by the modified metrics we argued that the modified version of GTD metric are most suitable metric structures for studying the underlying thermodynamic behavior of a black hole system.
\end{abstract}

\section{Introduction}
\label{sec: intro}
Black hole (BH) thermodynamics is a framework to study the thermodynamic properties of black hole system. Basic formulation of BH thermodynamics shows that the thermodynamical properties of the BH system is highly dependent on the geometry of the space-time created by the black hole \cite{Padmanabhan_2002,Jacobson_1995,Wald_1993}. Based on the space-time geometry of the black hole one can easily define the macroscopic thermodynamic properties such as internal energy, entropy etc. of the black hole based on four laws of black hole thermodynamics \cite{Bardeen_1973_4_law}. But to study the microscopic behavior of the constituent particle one need a different tool which can derive the microscopic properties from macroscopic properties; like a reverse way of statistical mechanics. Thermodynamic geometry or information geometry is one of such formalism. Thermodynamic geometry mainly based on defining a metric structure in thermodynamic phase space and applying the tool of general relativity on that thermodynamic phase space to investigate microscopic thermodynamic behavior \cite{Janyszek_1989,Aman_2006,Shen_2007,Brandner_2020,Xu_2020,Mirza_2011}. Most famous approach of thermodynamic geometry is the Weinhold's approach, in which hessian of mass or internal energy of the thermodynamic system (black hole) is treated as the metric tensor in the thermodynamical phase space \cite{Weinhold_1975}. On a similar way, by investigating the thermodynamic fluctuation theory Ruppeiner defined another metric in thermodynamic phase space as negative hessian of the entropy function \cite{Ruppeiner_1979}. In thermodynamic geometry both Weinhold metric and Ruppeiner metric has been extensively used in investigating various properties of a thermodynamic system such as thermodynamic length, Riemannian structure, thermodynamic curvature etc. \cite{Salamon_1980,Salamon_1984,Salamon_1985,Ruppeiner_1979,HERNANDEZ_1998,Shen_2007,Janke_2004,Johnston_2003,Santoro_2005}. But due to the lake of Legendre invariant nature, both Weinhold and Ruppeiner metric are not suitable for studying all the thermodynamic behavior of a system, e.g. behavior of the same thermodynamic system in different ensemble (for more details see \cite{Gogoi_2023}). To overcome these short coming H. Quevedo in 2007 \cite{Quevedo_2007} proposed an alternative method known as Geometrothermodynamics (GTD), which allows to defined a metric tensor in the thermodynamic phase space, incorporating the condition that the metric need to be invariant under Legendre transformation. Apart from the application of GTD approach in normal thermodynamic system such as ideal gas or van-dar walls gas \cite{Quevedo_2007,Quevedo_2022}, this approach is popularly used in investigating thermodynamic behavior of various BH system \cite{Alvarez_2008,Quevedo_2008a,Quevedo_2009_BTZ,Quovedo_2009_2D,Vazquez_2010,Quevedo_2008b}.

Bardeen black hole is the first regular BH solution of in general relativity proposed by Bardeen in 1968, in which black hole solution is interpreted as gravitationally collapsed magnetic monopole in non-linear theory of electrodynamics \cite{Bardeen_1968}. The distinguishable characteristic of regular black hole is the non-existence of central singularity unlike other class of BH system. But, it is important to note that apart from curvature singularities, BH thermodynamics demands the possible presence of a second types of singularities called thermodynamic singularity or Davis singularity. Inspired from classical theory of thermodynamics, one can define specific heat of black hole system at constant hair and it can be show that for some BH system like Kerr family of black hole the specific heat diverges at some thermodynamic space-time point generally referred as Devis point \cite{Davies_1978}. These Devis points defines the 2nd types of singularities so called thermodynamic singularities of a BH system. Although there is no curvature singularities in Bardeen regular BH, but it have thermodynamic singularities; which makes Bardeen BH an interesting thermodynamic BH system to study. Thermodynamics and Geometrothermodynamics (GTD) behavior of Bardeen regular BH is also reported in literature \cite{Quevedo_2024,Rodrigues_2022}. Although various literature have explored the thermodynamic curvature and phase transition of regular black hole \cite{Quovedo_2009_2D,Quevedo_2024,Rodrigues_2022}, a very few of them have concentrated on the behavior of thermodynamic geodesic in such system. In this work we have investigated the behavior of thermodynamic geodesic of Bardeen BH near criticality in three different GTD metric. Based on the behavior of the thermodynamic geodesic defined by conventional GTD metric, we have also modified the GTD metric in such a way that it can reflect most of the thermodynamic properties of the black hole system. The content of the paper is arranged as follows. In section \ref{sec:GTD}, we discuss fundamentals of GTD approach, in section \ref{sec: Thermodynamics_BH}, we discuss basic thermodynamics of Bardeen regular black hole, in section \ref{sec: Geodesic}, we investigate the behavior of thermodynamic geodesics of Bardeen regular BH defined by conventional GTD metric, in section \ref{sec: mod_metric}, we modified the conventional GTD metric and investigate the behavior of thermodynamic geodesics in the thermodynamic space defined by the modified metrics, in section \ref{sec: Results}, we discuss our results and explore behavior of thermodynamic curvature, and finally in section \ref{sec: conclusion}, we briefly summarized our result and mention the future aspects of this study.

\section{\label{sec:GTD}Geometrothermodynamics}

\noindent Thermodynamics system in Geometrothermodynamics (GTD) approach is represented mainly by three parameters, viz. $Z^{A} = \{\Phi, E^a, I^a\}$, with $a={1,2,3,...n}$, where $n$ is the thermodynamic dimension of the system \cite{Quevedo_2007}. Where $\Phi$ is the thermodynamic potential, $E^a$ is any extensive variable, and $I^a$ is corresponding intensive variable with $I^a = \frac{\partial \Phi}{\partial E^a}$. Main advantage of GTD formalism is that GTD metric define a thermodynamic system with (2n+1) dimensional manifold, which remains invariant under Legendre transformations of the co-ordinate $Z^A$.

H. Quevedo argued that black-hole system should be treated as quasi -homogeneous system \cite{Quevedo_2019_QBH}. For a thermodynamic system with $n$ degrees of freedom (dof), the fundamental equation can be written as 
\begin{equation}
    \Phi = \Phi(E^{a})
\end{equation}
If $\Phi$ is a homogeneous function,
\begin{equation}
    \Phi(\lambda^{\beta_a} E^a) = \lambda^{\beta_\Phi} \Phi(E^a)
    \label{eq:homoginity}
\end{equation}
Where $\beta_a$ is the degree of homogeneity with $\beta_a > 0$. For quasi-homogeneous system these $\beta$s can be differ from 1.

For a quasi-homogeneous system, the three different GTD metrics $\mathcal{G}^I$, $\mathcal{G}^{II}$, and $\mathcal{G}^{III}$ can be defined as (for more details see \cite{Quevedo_2019_QBH,Quevedo_2024})

\begin{align}
\mathcal{G}^I &= \sum_{a,b,c=1}^{n} \left( \beta_c E^c \frac{\partial \Phi}{\partial E^c} \right) \frac{\partial^2 \Phi}{\partial E^a \partial E^b} \, dE^a \, dE^b \label{eq:gI}\\
\mathcal{G}^{II} &= \sum_{a,b,c,d=1}^{n} \left( \beta_c E^c \frac{\partial \Phi}{\partial E^c} \right) \eta^d_a \frac{\partial^2 \Phi}{\partial E^b \partial E^d} \, dE^a \, dE^b \label{eq:gII}\\
\mathcal{G}^{III} &= \sum_{a,b=1}^{n} \left( \beta_a E^a \frac{\partial \Phi}{\partial E^a} \right) \frac{\partial^2 \Phi}{\partial E^a \partial E^b} \, dE^a \, dE^b \label{eq:gIII}
\end{align}

Where $\eta_a^d$ is a diagonal matrix of $n$ dimension with matrix elements diag(-1,1,...,1). These three different metrics can be used to investigate the thermodynamic behavior of black hole system under quasi-homogeneous scenario.

\section{\label{sec: Thermodynamics_BH}Thermodynamics of Bardeen regular Black Hole}
Under spherically symmetric approximation, regular solution of Bardeen black hole can be expressed as \cite{Bardeen_1968}
\begin{equation}
    ds^2 = -f(r)dt^2 +f(r)^{-1} dr^2 + r^2\, d\Omega^2
\end{equation}
Where 
\begin{equation}
    f(r) = 1 - \frac{2Mr^2}{(r^2+g^2)^{\frac{3}{2}}}
\end{equation}
with $M$ - Mass and $g$ - magnetic charge. \\
Event horizon of the black hole can be obtained by $f(r_h) = 0$, where $r_h$ is the radius of event horizon \cite{Capela_2012}. $f(r_h) = 0$ gives,
\begin{equation}
    M = \frac{(r_h^2+g^2)^{3/2}}{2r_h^2}
    \label{eq:M_r}
\end{equation}

Black-Hole thermodynamics (Bekenstein-Hawking area entropy relation) demands entropy to be
\begin{equation}
    S = \pi r_h^2
\end{equation}
From this relation lets rescale entropy as 
\begin{equation}
    s = \frac{S}{\pi} = r_h^2
    \label{eq:s_r}
\end{equation}
From equation (\ref{eq:M_r}) and (\ref{eq:s_r}) mass-entropy relation for Bardeen regular black hole can be written as 
\begin{equation}
    M = \frac{(s+g^2)^{3/2}}{2s}
    \label{eq:fundamental}
\end{equation}

Equation (\ref{eq:fundamental}) fundamental thermodynamic equation for Bardeen regular black hole with potential $M$. Clearly, Bardeen Black hole is a thermodynamic system with 2 dof, with extensive variable as entropy $s$ and magnetic charge $g$. The corresponding intensive variable can be defined as 
\begin{align}
    t = \left(\frac{\partial M}{\partial s} \right)_g = \frac{(s-2g^2)\sqrt{s+g^2}}{4s^2} \\
    \phi = \left(\frac{\partial M}{\partial g} \right)_s = \frac{3g\sqrt{s+g^2}}{2s}
\end{align}

Where $t$, the entropy derivative of thermodynamic potential $M$ is the Hawking temperature \cite{Hawking_1974}, and $\phi$ is the magnetic potential corresponding to the magnetic charge $g$.

To study the criticality and phase transition, it is important to define specific heat of the black hole system. Specific heat at constant magnetic charge $g$ can be defined as 
\begin{align}
    c_g &= t \left( \frac{\partial s}{\partial t} \right)_g = \frac{\left( \frac{\partial M}{\partial s} \right)_g}{\left( \frac{\partial^2 M}{\partial s^2} \right)_g} = \frac{\left( \frac{\partial M}{\partial s} \right)_g}{\left( \frac{\partial t}{\partial s} \right)_g} \notag \\
    c_g &=  \frac{2s(-2g^2 + s)(g^2 + s)}{8g^4 + 4g^2 s - s^2}
\end{align}

In 2nd order phase transition, specific heat  diverges at the transition points. So, following Davies argument, transition points for the 2nd order phase-transition can be obtained by equating $\frac{1}{c_g} = 0$ \cite{Davies_1978}, which gives 

\begin{equation}
    s = 2(1+\sqrt{3})g^2 \label{eq:spinodal}
\end{equation}

\begin{figure}
    \centering
    \includegraphics[width=\linewidth]{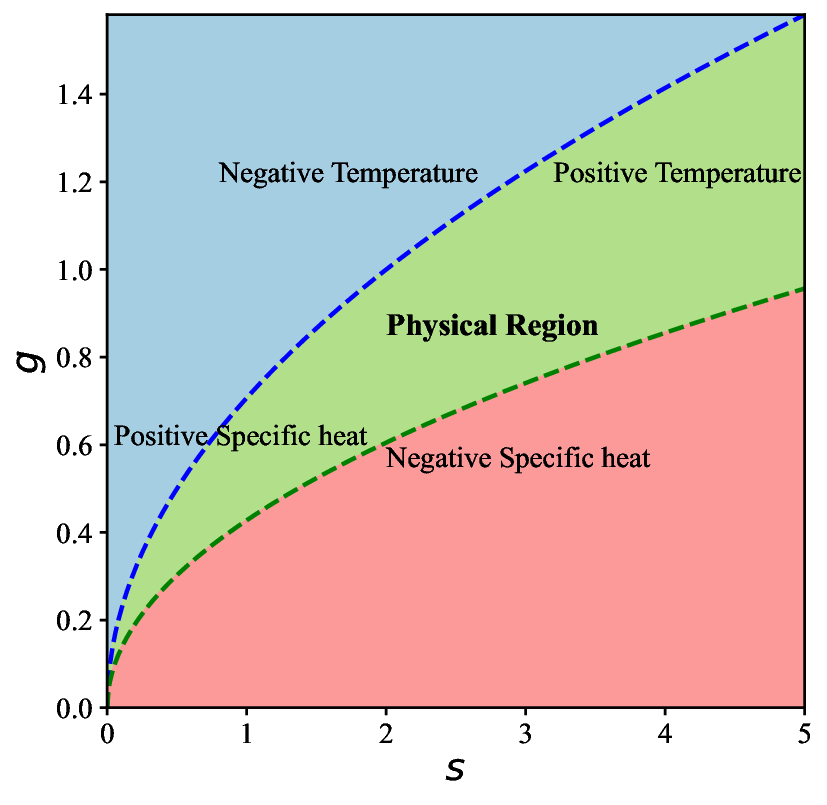}
    \caption{Locus of Davis points and temperature vanishing points of Bardeen regular black hole in $s-g$ plane. Green dashed line represents the spinodal line (Devis points)  which separate positive specific heat region from negative specific heat region, while blue dashed line represent the temperature vanishing line,  which separate positive temperature region from negative temperature region. Blue colored region represents the negative temperature region, red colored region represents the negative specific heat region, while green colored region represents the Physical region (PR); the region with positive specific heat and positive temperature.}
    \label{fig:physical_region}
\end{figure}

This equation represent a curve in $s-g$ plane called 
 ``the spinodal curve", which separate positive specific heat region to negative specific heat region. Apart from spinodal curve, there exist one more boundary in black hole thermodynamic system, ``the temperature vanishing curve", which separate positive temperature region to negative temperature region. Temperature vanishing curve can be obtained by the equation $t=0$ as,
\begin{equation}
    s = 2g^2
    \label{eq:t=0}
\end{equation}
So physical region (PR) of the black hole system in the thermodynamic plane is the region enclosed by both spinodal curve and temperature vanishing curve, with positive specific heat and positive temperature (see fig \ref{fig:physical_region}). In this work we will explore the behavior thermodynamic geodesic of Bardeen regular BH in the physical region considering all $\mathcal{G}^I$, $\mathcal{G}^{II}$, and $\mathcal{G}^{III}$ metric as represented by equation (\ref{eq:gI}), (\ref{eq:gII}), and (\ref{eq:gIII}). Before investigating thermodynamic geodesic, it is also important to calculate the degrees of homogeneity $\beta^a$. Substituting equation (\ref{eq:homoginity}) in the fundamental equation (\ref{eq:fundamental}), it can be easily shown that
\begin{equation}
    \beta_s = 2\beta_g = 2\beta_M
\end{equation}
So, for our purpose, we can choose degree of homogeneities as $(\beta_s,\beta_g,\beta_M) =(2,1,1)$.

\section{\label{sec: Geodesic}Thermodynamic Geodesics}
For any metric tensor $g_{\mu \nu}$, one can calculate the extrema of the integral $\int_1^2 d\lambda \sqrt{g_{\mu \nu} \dot{x}^\mu \dot{x}^\nu}$ in the Riemannian spacetime and use variational principle to obtain the geodesic equation as  
\begin{equation}
    \ddot{x}^\mu + \Gamma^\mu_{\nu \rho} \dot{x}^{\nu}\dot{x}^\rho = 0
\end{equation}
Where $\lambda$ is the affine parameter, derivative with respect to $\lambda$ is represented by dot(s). To obtain the geodesic equation one can also define a Lagrangian in $d$ dimensional thermodynamic space as $\mathscr{L} = g_{\mu \nu} \dot{x}^{\mu}\dot{x}^\nu$, and geodesic equations become
\begin{equation}
    \frac{\partial}{\partial \lambda}\left(\frac{\partial \mathscr{L}}{\partial \dot{x}^{\mu}}\right) -\frac{\partial \mathscr{L}}{\partial x^\mu} = 0
    \label{eq:geodesic_L}
\end{equation}
Where $\mu = (1,2,...d)$, $d$ is the thermodynamic dimension of the system. \\
It is clear from fundamental equation (\ref{eq:fundamental})  that Bardeen regular black hole is a two dimensional thermodynamic system with thermodynamic co-ordinate $(s,g)$. So, in this work we will explore geodesic behavior of Bardeen regular BH in three metrics defined by equation (\ref{eq:gI},\ref{eq:gII},\ref{eq:gIII}) in $(s,g)$ co-ordinate system.
\subsection{Geodesics in $\mathcal{G}^I$ metric}
Substituting the thermodynamic potential, and extensive variables, equation (\ref{eq:gI}) can be reduces to
\begin{equation}
    ds^2_I = \mathcal{G}^I = (2st+g\phi)\, \left( \frac{\partial^2 M}{\partial s^2} ds^2 + 2\frac{\partial ^2M}{\partial s \partial g}ds dg +\frac{\partial ^2 M}{\partial g^2}dg^2\right)
\end{equation}
So metric element can be expressed as

\begin{equation}
\mathcal{G}^I = \begin{pmatrix}
    M_{ss} & M_{sg} \\
    M_{gs} & M_{gg}
\end{pmatrix}
\end{equation}
With $M_{ss} = -\frac{\left(g^2+s\right) \left(-8 g^4-4 g^2 s+s^2\right)}{16 s^4}$,\\ $M_{sg} = M_{gs} = \left(\frac{3 g}{4 s \sqrt{g^2+s}}-\frac{3 g \sqrt{g^2+s}}{2 s^2}\right) $\\$\left(2 s \left(\frac{3 \sqrt{g^2+s}}{4 s}-\frac{\left(g^2+s\right)^{3/2}}{2 s^2}\right)+\frac{3 g^2 \sqrt{g^2+s}}{2 s}\right)$ and \\
$M_{gg} = \frac{3 \left(g^2+s\right) \left(2 g^2+s\right)}{4 s^2}$.\\
The geodesic equation corresponding to this metric structure can be obtained from equation (\ref{eq:geodesic_L}) as
\begin{multline}
    s' \left(\frac{6 g^5 g'}{s^4}+\frac{6 g^3 g'}{s^3}+\frac{3 g g'}{4 s^2}\right) -\frac{3 g^4 \left(g g''+3 g'^2\right)}{2 s^3} \\
    -\frac{9 g^2 \left(g g''+2 g'^2\right)}{4 s^2}-\frac{3 \left(g g''+g'^2\right)}{4 s}+\\
    \left(\frac{g^6}{s^4}+\frac{3 g^4}{2 s^3}+\frac{3 g^2}{8 s^2}-\frac{1}{8 s}\right) s'' \\
    +\left(-\frac{2 g^6}{s^5}-\frac{9 g^4}{4 s^4}-\frac{3 g^2}{8 s^3}+\frac{1}{16 s^2}\right) s'^2 = 0
    \label{eq:g1_geo1}
\end{multline}
and
\begin{multline}
    \left(\frac{3 g^4}{s^2}+\frac{9 g^2}{2 s}+\frac{3}{2}\right) g''+ g' \left(-\frac{6 g^4 s'}{s^3}-\frac{9 g^2 s'}{2 s^2}\right)+ \\\left(\frac{6 g^3}{s^2}+ \frac{9 g}{2 s}\right) g'^2
    +\frac{3 g^5 \left(s'^2-s s''\right)}{2 s^4}+ \\
    \frac{g^3 \left(6 s'^2-9 s s''\right)}{4 s^3}+
    \frac{3 g \left(s'^2-2 s s''\right)}{8 s^2} =0
    \label{eq:g1_geo2}
\end{multline}

\begin{figure}
    \centering
    \includegraphics[width=\linewidth]{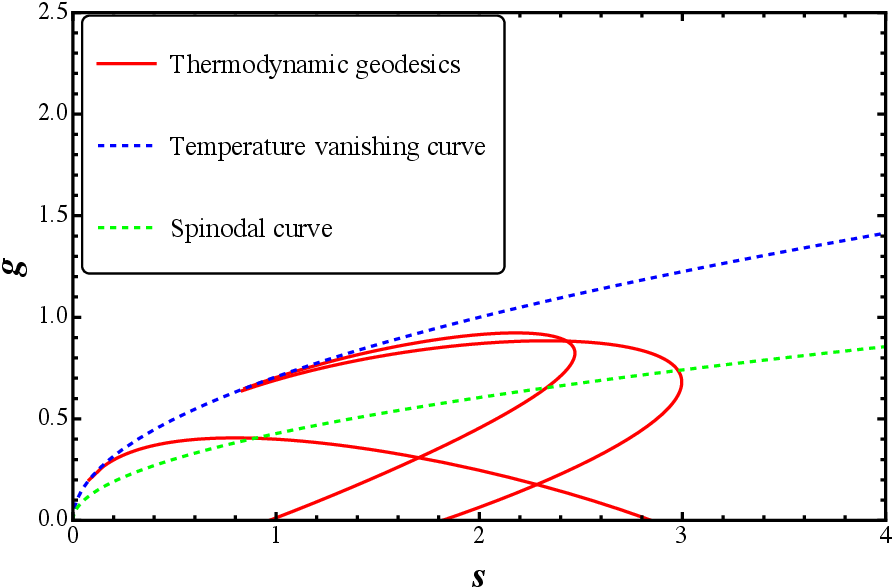}
    \caption{Thermodynamic geodesic of Bardeen regular black hole in thermodynamic $s-g$ space defined by the metric $\mathcal{G}^I$.}
    \label{fig:g1 geosesic}
\end{figure}

Derivative with respect to $\lambda$ is represented by prime(s). We solve the geodesic equation (\ref{eq:g1_geo1}) and (\ref{eq:g1_geo2}) numerically for different boundary conditions and plotted the different geodesics in $s-g$ plane. It is observed that all the geodesic in $s-g$ plane exhibit an incompleteness behavior near the temperature vanishing curve, and no thermodynamic geodesic enter from positive temperature region to negative temperature region and vice versa. But thermodynamic geodesic defined by $\mathcal{G}^I$ metric does not show any turn around behavior or incompleteness near the spinodal line. These thermodynamic geodesic can cross the spinodal line and enter the negative specific heat region. Hence geodesic defined by metric $\mathcal{G}^I$ does not confined in the physical region or single phase as expected. Fig \ref{fig:g1 geosesic} shows the behavior of thermodynamic geodesic in $s-g$ phase space defined by the metric $g^I$.

\subsection{Geodesic in $\mathcal{G}^{II}$ metric}
Line element in $\mathcal{G}^{II}$ space is defined by
\begin{equation}
    ds^2_{II} = g^{II} = (2st+g\phi)\, \left(- \frac{\partial^2 M}{\partial s^2} ds^2  +\frac{\partial ^2 M}{\partial g^2}dg^2\right)
    \label{eq:g2}
\end{equation}
So metric element become
\begin{equation}
\mathcal{G}^{II} = \begin{pmatrix}
    M_{ss} & 0 \\
    0 & M_{gg}
\end{pmatrix}
\label{eq:g2 metric}
\end{equation}
With $M_{ss}= \frac{\left(g^2+s\right) \left(-8 g^4-4 g^2 s+s^2\right)}{16 s^4}$ and \\
$M_{gg} = \frac{3 \left(g^2+s\right) \left(2 g^2+s\right)}{4 s^2}$.
With this geodesic equation in $\mathcal{G}^{II}$ space can be obtained from equation (\ref{eq:geodesic_L}) as 
\begin{multline}
    s' \left(-\frac{6 g^5 g'}{s^4}-\frac{6 g^3 g'}{s^3}-\frac{3 g g'}{4 s^2}\right) + \frac{3 g^4 g'^2}{s^3}+\frac{9 g^2 g'^2}{4 s^2}+ \\
    \left(-\frac{g^6}{s^4}-\frac{3 g^4}{2 s^3}-\frac{3 g^2}{8 s^2}+ \frac{1}{8 s}\right) s''+ \\
    \left(\frac{2 g^6}{s^5}+\frac{9 g^4}{4 s^4}+\frac{3 g^2}{8 s^3}-\frac{1}{16 s^2}\right) s'^2 =0
    \label{eq:gII_geo1}
\end{multline}
and
\begin{multline}
    \left(\frac{3 g^4}{s^2}+\frac{9 g^2}{2 s}+\frac{3}{2}\right) g''+ g' \left(-\frac{6 g^4 s'}{s^3}-\frac{9 g^2 s'}{2 s^2}\right)+\\
    \left(\frac{6 g^3}{s^2}+\frac{9 g}{2 s}\right) g'^2+\frac{3 g^5 s'^2}{s^4}+ \frac{3 g^3 s'^2}{s^3}+\frac{3 g s'^2}{8 s^2} = 0
    \label{eq:gII_geo2}
\end{multline}

\begin{figure}
    \centering
    \includegraphics[width=\linewidth]{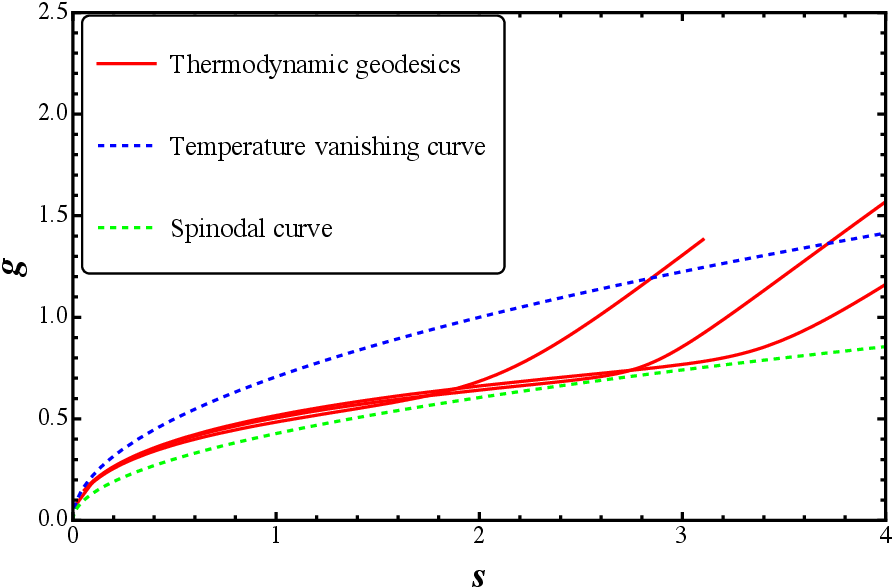}
    \caption{Thermodynamic geodesic of Bardeen regular black hole in thermodynamic $s-g$ space defined by the metric $\mathcal{G}^{II}$.}
    \label{fig:g2 geodesic}
\end{figure}

We solve the geodesic equation (\ref{eq:gII_geo1}) and (\ref{eq:gII_geo2}) numerically and plotted in $s-g$ plane. It is observed that unlike $\mathcal{G}^{I}$ metric space these geodesic does not show a incomplete behavior near the temperature vanishing line rather they show a turn around behavior near the spinodal line. In phase space defined by metric $\mathcal{G}^{II}$, it is observe that no geodesic enter from positive specific heat region to negative specific heat region and vice versa. But geodesic defined by $\mathcal{G}^{II}$ metric can easily cross the temperature vanishing curve and enter in the negative temperature region. Hence geodesic in $\mathcal{G}^{II}$ metric formalism also does not confine in the physical region. Fig \ref{fig:g2 geodesic} shows the behavior of thermodynamic geodesic in physical region in the thermodynamic space defined by $\mathcal{G}^{II}$ metric.

\subsection{Geodesic in $\mathcal{G}^{III}$ metric}
Line element in $\mathcal{G}^{III}$ space is defined by
\begin{multline}
    ds^2_{III} = \mathcal{G}^{III} = 2st\,  \frac{\partial^2 M}{\partial s^2} ds^2 + 2st\frac{\partial ^2M}{\partial s \partial g}ds dg + \\
    g\phi\frac{\partial ^2M}{\partial g \partial s}dg ds+g\phi \frac{\partial ^2 M}{\partial g^2}dg^2
\end{multline}

So metric tensor become
\begin{equation}
\mathcal{G}^{III} = \begin{pmatrix}
    M_{ss} & M_{sg} \\
    M_{gs} & M_{gg}
\end{pmatrix}
\label{eq:g3 metric}
\end{equation}
with $M_{ss}= -\frac{16 g^6-6 g^2 s^2+s^3}{16 s^4}$,\\ $M_{sg} = M_{gs}=-\frac{3 g \left(g^2+s\right) \left(2 g^2+s\right)}{8 s^3}$, and $M_{gg} = \frac{9 g^2 \left(2 g^2+s\right)}{4 s^2}$.
With the thermodynamic space defined by equation (\ref{eq:g3 metric}), geodesic equation becomes,
\begin{multline}
    s' \left(\frac{3 g g'}{2 s^2}-\frac{12 g^5 g'}{s^4}\right)+\frac{3 g^4 \left(g'^2-g g''\right)}{2 s^3}-\frac{9 g^2 \left(g g''+2 g'^2\right)}{4 s^2} \\
    -\frac{3 \left(g g''+g'^2\right)}{4 s}+ \left(-\frac{2 g^6}{s^4}+\frac{3 g^2}{4 s^2}-\frac{1}{8 s}\right) s''+ \\
    \left(\frac{4 g^6}{s^5}-\frac{3 g^2}{4 s^3}+\frac{1}{16 s^2}\right) s'^2 = 0
    \label{eq:g3_geo1}
\end{multline}
and 
\begin{multline}
    \left(-\frac{3 g^5}{2 s^3}-\frac{9 g^3}{4 s^2}-\frac{3 g}{4 s}\right)g''+g' \left(\frac{3 g s'}{2 s^2}-\frac{12 g^5 s'}{s^4}\right) \\
    +\left(\frac{3 g^4}{2 s^3}-\frac{9 g^2}{2 s^2}-\frac{3}{4 s}\right) g'^2+\frac{g^6 \left(4 s'^2-2 s s''\right)}{s^5} \\
    +\frac{3 g^2 \left(s s''-s'^2\right)}{4 s^3}+\frac{s'^2-2 s s''}{16 s^2} = 0
    \label{eq:g3_geo2}
\end{multline}

\begin{figure}
    \centering
    \includegraphics[width=\linewidth]{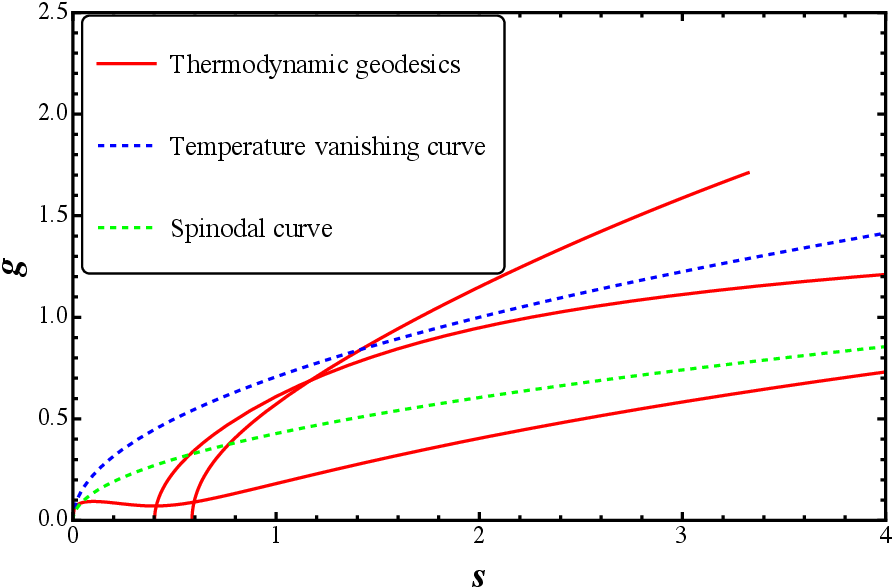}
    \caption{Thermodynamic geodesic of Bardeen regular black hole in thermodynamic $s-g$ space defined by the metric $\mathcal{G}^{III}$.}
    \label{fig:g3 geodesic}
\end{figure}

We solve equation (\ref{eq:g3_geo1}) and (\ref{eq:g3_geo2}) numerically and plotted in $s-g$ plane. It is very interesting to note that geodesic defined by $\mathcal{G}^{III}$ metric cross both spinodal and temperature vanishing line, and hence these geodesic does not confined in the physical region at all. The behavior of the geodesics in thermodynamic space defined by $\mathcal{G}^{III}$ metric is shown in fig \ref{fig:g3 geodesic}. \\

It appears that neither of the metric $\mathcal{G}^{I}$, $\mathcal{G}^{II}$, or $\mathcal{G}^{III}$ is appropriate to describe the complete thermodynamic behavior of the system, as none of this metric contain complete information about the both temperature and specific heat of the system. Below we redefine conventional GTD metrics and investigate the behavior of the system in space defined by the modified metrics.

\section{\label{sec: mod_metric}Modified GTD metrics}

Non-confinement of thermodynamic geodesic in the physical region defined by conventional GTD metrics suggest that it is necessary to make some modifications in the metric structure of the conventional GTD metrics. While modifying GTD metrics it is important to keep in mind that Legendre invariant nature of GTD metrics have to be preserve in its metric structure. H. Quevedo \cite{Quevedo_2007} in 2007 shows that simplest modification to make thermodynamic metric such as Weinhold metric ($\mathcal{G}_w$) Legendre invariant is

\begin{equation}
    \mathcal{G}_{GTD} = M\mathcal{G}_w = M \frac{\partial^2M}{\partial E^a \partial E^b} dE^a dE^b
\end{equation}

From Euler identity, the potential $M$ can be expressed as summation of all the multiplicative pairs of extensive-intensive variables required to describe the thermodynamic system, i.e. $M=\displaystyle \sum_{c} \Big(E_c \frac{\partial M}{\partial E_c} \Big)$. With this, Legendre invariant version of Weinhold metric become
\begin{equation}
    \mathcal{G}_{GTD} = \sum_{c} \Big(E_c \frac{\partial M}{\partial E_c} \Big) \frac{\partial^2M}{\partial E^a \partial E^b} dE^a dE^b
\end{equation}

With the introduction of quasi-homogenous parameters ($\beta$s) this equation exactly mimic equation for $\mathcal{G}^I$ metric (equation \ref{eq:gI}). But, in this work it is observed that introduction of all the extensive-intensive pair terms of the thermodynamic potential $M$ (from Euler identity) results in non-confinement of geodesics in a single phase. So here we modify these metrics in such a way that only one extensive-intensive pair such as entropy-temperature pair appears in the metric elements. It is straightforward to show that introduction of only one extensive-intensive multiplicative pair as a coefficient of mass hessian is enough to preserve the Legendre invariant nature of GTD metrics (for more details see \cite{Quevedo_2007}). GTD metric with only entropy-temperature pair is also reported in literature \cite{Gogoi_2023,Rani_2024}. With taking care of these aspects, the conventional GTD metrics can be modified as

\begin{align}
    \mathcal{G}^{II}_{mod} &= \sum_{a,b,c,d=1}^{n} \left( \beta_c E^c \frac{\partial \Phi}{\partial E^c} \delta^c_1 \right) \eta^d_a \frac{\partial^2 \Phi}{\partial E^b \partial E^d} \, dE^a \, dE^b \label{eq:gII_mod}\\
    \mathcal{G}^{III}_{mod} &= \sum_{a,b=1}^{n} \left( \beta_a E^a \frac{\partial \Phi}{\partial E^a} \right) \delta^b_a\frac{\partial^2 \Phi}{\partial E^a \partial E^b} \, dE^a \, dE^b \label{eq:gIII_mod}
\end{align}

Where $\delta$s are Kronecker delta of $n$ dimension, and $E^1 =s$. In the modified metrics we introduce a term $\delta_1^c$ in $\mathcal{G}^{II}$ metric structure. This term ensures that instead of the whole potential $M$ only the entropy-temperature ($st$) pair term will appear in the GTD metric elements. Similarly in $\mathcal{G}^{III}$ metric structure it is observe that only off-diagonal elements of the metric contain the potential term $M$, while diagonal elements contain only one extensive-intensive pair. So preserving this metric structure, we introduce a term $\delta_a^b$ in the equation of $\mathcal{G}^{III}$ metric to ensure that only diagonal term of the metric will contribute enforcing the off-diagonal terms to be zero. Note that one possible modification of $\mathcal{G}^I$ metric is

\begin{equation}
    \mathcal{G}^I_{mod} = \sum_{a,b,c=1}^{n} \left( \beta_c E^c \frac{\partial \Phi}{\partial E^c} \delta^c_1  \right) \delta^b_a \frac{\partial^2 \Phi}{\partial E^a \partial E^b} \, dE^a \, dE^b \nonumber
    \label{eq:gI_mod}
\end{equation}

    Since this metric structure is very much similar to $\mathcal{G}^{II}_{mod}$ metric, so we are considering only $\mathcal{G}^{II}_{mod}$ and $\mathcal{G}^{III}_{mod}$ metric structure in this work. 

In this section we will investigate the properties of thermodynamic geodesic in the thermodynamic space defined by the metrics (\ref{eq:gII_mod}) and (\ref{eq:gIII_mod}).

\subsection{Geodesic in $\mathcal{G}^{II}_{mod}$ metric}
Line element is $\mathcal{G}^{II}_{mod}$ space become
\begin{equation}
    ds^2_{II'} = \mathcal{G}^{II}_{mod} = 2st\, \left(- \frac{\partial^2 M}{\partial s^2} ds^2  +\frac{\partial ^2 M}{\partial g^2}dg^2\right)
    \label{eq:g2_mod}
\end{equation}

Similar metric structure also used in studying thermodynamic behavior of Karr family Black hole and Dynonic Black hole \cite{Gogoi_2023}, as well as in regular black hole system \cite{Rani_2024}.
Metric tensor can be expressed as 
\begin{equation}
\mathcal{G}^{II} = \begin{pmatrix}
    M_{ss} & 0 \\
    0 & M_{gg}
\end{pmatrix}
\end{equation}

With $M_{ss} = \frac{16 g^6-6 g^2 s^2+s^3}{16 s^4}$, and $M_{gg} = \frac{3}{4}-\frac{3 g^4}{s^2}$. Geodesic equation defined by this metric can be expressed as 
\begin{multline}
    s' \left(\frac{12 g^5 g'}{s^4}-\frac{3 g g'}{2 s^2}\right)-\frac{6 g^4 g'^2}{s^3}+\left(\frac{2 g^6}{s^4}-\frac{3 g^2}{4 s^2}+\frac{1}{8 s}\right) s''\\+ 
    \left(-\frac{4 g^6}{s^5}+\frac{3 g^2}{4 s^3}-\frac{1}{16 s^2}\right) s'^2 = 0
    \label{eq:g2_mod_geo1}
\end{multline}

and

\begin{multline}
    \left(\frac{3}{2}-\frac{6 g^4}{s^2}\right) g''+\frac{12 g^4 g' s'}{s^3}-\frac{12 g^3 g'^2}{s^2} -\frac{6 g^5 s'^2}{s^4}+\frac{3 g s'^2}{4 s^2} =0
    \label{eq:g2_mod_geo2}
\end{multline}

\begin{figure}
    \centering
    \includegraphics[width=\linewidth]{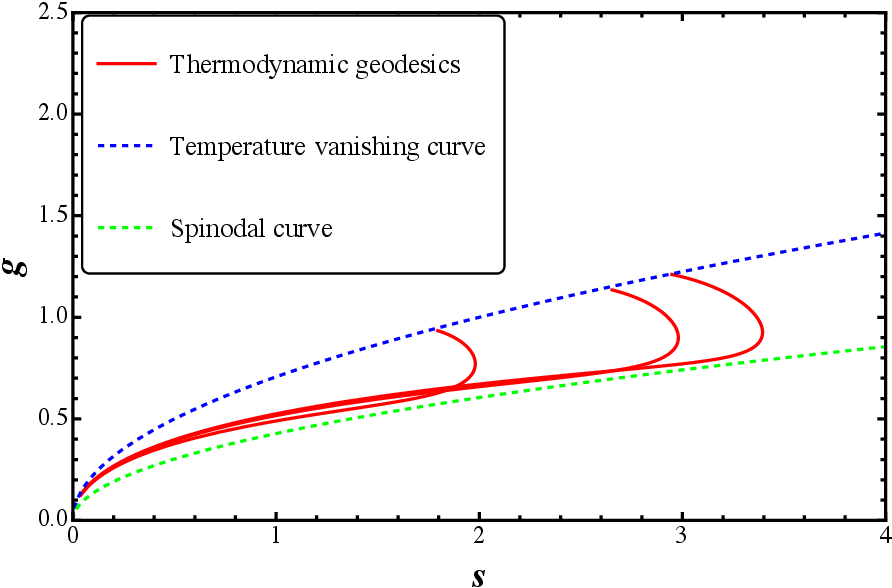}
    \caption{Thermodynamic geodesic of Bardeen regular black hole in thermodynamic $s-g$ space defined by the metric $\mathcal{G}^{II}_{mod}$}
    \label{fig:g2_mod geodesic}
\end{figure}

We solve geodesic equation (\ref{eq:g2_mod_geo1}) and (\ref{eq:g2_mod_geo2}) numerically and plotted in $(s-g)$ plane (fig \ref{fig:g2_mod geodesic}). It is very interesting to observe that unlike $\mathcal{G}^{I}$, $\mathcal{G}^{II}$, and $\mathcal{G}^{III}$ metric,
these geodesic cross neither spinodal line nor temperature vanishing line. Thermodynamic geodesic defined by $\mathcal{G}^{II}_{mod}$ matrix exhibit a turn around behavior near the spinodal line and shows incompleteness towards temperature vanishing line, and hence these geodesic confined in the physical region only. 

\subsection{Geodesic in $\mathcal{G}^{III}_{mod}$ metric}

Line element in thermodynamic space defined by $\mathcal{G}^{III}_{mod}$ metric can be expressed as 

\begin{multline}
    ds^2_{III'} = \mathcal{G}^{III}_{mod} = 2st\,  \frac{\partial^2 M}{\partial s^2} ds^2 +g\phi \frac{\partial ^2 M}{\partial g^2}dg^2
\end{multline}

\begin{figure}
    \centering
    \includegraphics[width=\linewidth]{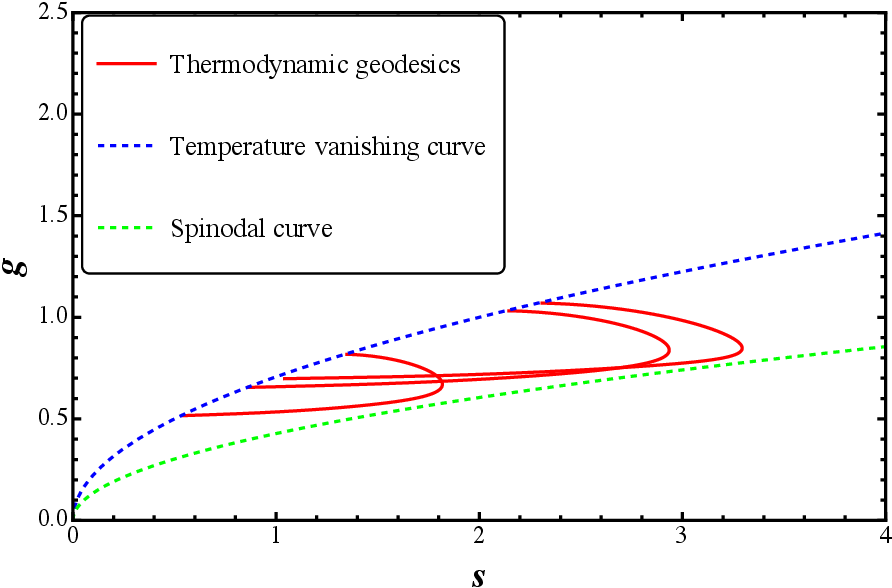}
    \caption{Thermodynamic geodesic of Bardeen regular black hole in thermodynamic $s-g$ space defined by the metric $\mathcal{G}^{III}_{mod}$}
    \label{fig:g3 mod geodesic}
\end{figure}

Similarly, metric tensor become
\begin{equation}
\mathcal{G}^{III} = \begin{pmatrix}
    M_{ss} & 0 \\
    0 & M_{gg}
\end{pmatrix}
\label{eq:g1 metric}
\end{equation}
With $M_{ss} =-\frac{16 g^6-6 g^2 s^2+s^3}{16 s^4}$, and $M_{gg} = \frac{9 g^2 \left(2 g^2+s\right)}{4 s^2}$. With this metric structure geodesic equation becomes,

\begin{multline}
    s' \left(\frac{3 g g'}{2 s^2}-\frac{12 g^5 g'}{s^4}\right)+\frac{9 g^4 g'^2}{s^3}+\frac{9 g^2 g'^2}{4 s^2}+\\\left(-\frac{2 g^6}{s^4}+\frac{3 g^2}{4 s^2}-\frac{1}{8 s}\right) s''+\left(\frac{4 g^6}{s^5}-\frac{3 g^2}{4 s^3}+\frac{1}{16 s^2}\right) s'^2 = 0
    \label{eq:g3_mod_geo1}
\end{multline}
and
\begin{multline}
    \left(\frac{9 g^4}{s^2}+\frac{9 g^2}{2 s}\right) g''+g' \left(-\frac{18 g^4 s'}{s^3}-\frac{9 g^2 s'}{2 s^2}\right)+\\
    \left(\frac{18 g^3}{s^2}+\frac{9 g}{2 s}\right) g'^2+\frac{6 g^5 s'^2}{s^4}-\frac{3 g s'^2}{4 s^2} = 0
\end{multline}
We solve these geodesic equation numerically and plotted in the thermodynamic plane defined by $\mathcal{G}^{III}_{mod}$ metric. It is observe that every geodesic shows a turn around behavior near the spinodal line and incompleteness near the temperature vanishing line. Similar to $\mathcal{G}^{II}_{\mathrm{mod}}$ metrics, geodesic defined by $\mathcal{G}^{III}_{\mathrm{mod}}$ metric is also confined in a single phase or physical region.  Fig \ref{fig:g3 mod geodesic} shows the behavior of thermodynamic geodesic in the physical region.

\section{\label{sec: Results}Results and Discussion}
In this work we computed geodesic of Bardeen regular BH in $s-g$ space with five different metric, three conventional GTD metric viz., $\mathcal{G}^{I}$, $\mathcal{G}^{II}$, and $\mathcal{G}^{III}$ and two modified GTD metric viz.,  $\mathcal{G}^{II}_{mod}$ and $\mathcal{G}^{III}_{mod}$. Among the conventional GTD metric, in $\mathcal{G}^{I}$ metric thermodynamic geodesic shows incompleteness behavior near the temperature vanishing line, $\mathcal{G}^{II}$ metric thermodynamic geodesic shows turn around behavior near the spinodal line, while in $\mathcal{G}^{III}$ metric geodesic cross both spinodal and temperature vanishing curve. But neither of these metric is able to reflect expected behavior of geodesic in physical region. Non-confinement of geodesic in a single phase indicates that these conventional three GTD metrics do not contain all information of thermodynamic behavior in a single metric tensor. To investigate the reason for inconsistency one should go a step backward. Although we expect incompleteness or turn around behavior near the temperature vanishing curve or spinodal curve, but in a true sense thermodynamic geodesic should show such behavior near those point at which thermodynamic curvature diverges. For a thermodynamic system with two dof thermodynamic curvature or Ricci scalar can be defined as 
\begin{multline}
     R = \\
     -\frac{1}{\sqrt{d}} \left[\frac{\partial}{\partial s} \left(  \frac{\partial_s M_{gg} -\partial_gM_{sg}}{\sqrt{d}} \right) + \frac{\partial}{\partial g} \left( \frac{\partial_g M_{ss} -\partial_s M_{sg}}{\sqrt{d}}\right) \right]
\end{multline}
 Where $d$ is the determinant of the metric tensor $d = M_{ss}M_{gg} - M_{sg}^2$, and $\partial_\mu  = \frac{\partial}{\partial \mu}$ with $\mu = (s,g)$.   

So, if at any locus of point the if curvature diverges, the thermodynamic geodesic should not cross the curvature diverging points, either by showing turn around behavior or by showing incompleteness depending on the nature of curvature. So to investigate nature of curvature we computed the thermodynamic curvature or Ricci scalar as a function of $(s-g)$ for all the three metric and investigate for possible singularities of the curvature scalar. Detailed formula of curvature scaler for all the metric structures are listed in appendix. Singularities of the thermodynamic curvature can be obtained by solving $\frac{1}{R} = 0.$ For $\mathcal{G}^{I}$ metric, the curvature diverges at
\begin{equation}
    g = \sqrt{\frac{s}{2}}
\end{equation}

In analogous to equation (\ref{eq:t=0}) one can easily identify this boundary as temperature vanishing line. But this curvature does not diverges near the Devis point (spinodal curve). So geodesic in $\mathcal{G}^{I}$ metric space can't sense the spinodal line but show incompleteness near the temperature vanishing line. Similarly curvature in $\mathcal{G}^{II}$ metric space diverges only at the Devis points hence geodesic defined by $\mathcal{G}^{II}$ metric can't sense the temperature vanishing line. Surprisingly curvature defined by $\mathcal{G}^{III}$ metric for positive $s$ and positive $g$ is complex and regular, although it have singularities and real value for negative values of magnetic charge. So any geodesic in positive $s-g$ plane will not be able to sense either spinodal or temperature vanishing curve, and hence will not confined in the physical region. Behavior of curvature scaler of all these metrics is summarized in table \ref{table: table1}. \\
As curvature defined by these metrics does not contain all the physical or thermodynamical information of the system, so in a true sense these metric are not appropriate to investigate thermodynamic behavior of a system. \\

On the other hand, curvature scaler of modified metric $\mathcal{G}^{II}_{mod}$ and $\mathcal{G}^{III}_{mod}$ diverges both at $c_g \rightarrow \infty$ and $t \rightarrow 0$. For these metrics physical region is bounded by the curve demanding both positive temperature and positive specific heat. Due to  this reason geodesic defined by these metric never enter the unphysical region and confined only in a single phase. These behavior make the modified metric suitable for investigating thermodynamic properties of black hole system. Behavior of thermodynamic curvature define by modified metric is summarized in table \ref{table: table1}.

\begin{table*}[ht]
    \centering
    \caption{Behavior of thermodynamic curvature and thermodynamic geodesics of Bardeen regular Black Hole with GTD and modified GTD metric.}
    \vspace{0.3em}
    \renewcommand{\arraystretch}{1.2}
    \setlength{\tabcolsep}{8pt}
    \begin{tabular}{lccc}
        \hline
        \textbf{Metric tensor} & \textbf{Curvature diverges at Davies points?} & \textbf{Curvature diverges at $T = 0$?} & \textbf{Geodesics confined in PR?} \\
        \hline
        $\mathcal{G}^I$              & No & Yes & No \\
        $\mathcal{G}^{II}$           & Yes & No & No \\
        $\mathcal{G}^{III}$          & No & No & No \\
        $\mathcal{G}^{II}_{\mathrm{mod}}$  & Yes & Yes & Yes \\
        $\mathcal{G}^{III}_{\mathrm{mod}}$ & Yes & Yes & Yes \\
        \hline
    \end{tabular}
    \label{table: table1}
\end{table*}

\section{\label{sec: conclusion}Conclusion}
In this work we investigate the behavior of thermodynamic geodesic in three different conventional GTD metrics, and also in two modified GTD metrics. Based on the nature of thermodynamic geodesic and thermodynamic curvature we argued that conventional GTD metric need slight modification to properly describe the thermodynamic behavior of a system. In the frame work of Bardeen regular black hole we show that thermodynamic geodesic defined by modified GTD metric confined within a single thermodynamic phase and they exhibit either turn around behavior or incompleteness near the boundary which separate physical region to unphysical region. With the modified metrics this behavior is expected to hold for other black hole systems as well. Such behavior of confinement of geodesic in a single phase for Karr family black hole and dyonic black hole is already reported in literature \cite{Gogoi_2023}. This study shows that it is an important area to explore relation between the metric structure and physical boundaries of thermodynamic parameters. It will be an important study to identify or formulate a metric tensor which contain most of the information about the boundaries of thermodynamic parameter in all ensembles. We will address this aspect in our future work.
\begin{acknowledgements}
We sincerely thank the anonymous reviewer for their valuable suggestions, which have significantly improved the quality of the manuscript.
\end{acknowledgements}

\appendix
\section{Thermodynamic curvature of Bardeen regular Black hole in various GTD as well as modified GTD metrics}

Thermodynamic curvature of Bardeen regular Black hole in different GTD and modified GTD metric are given by\\
% {\color{red}
\begin{multline}
  \mathcal{G}^{I} :  R = \frac{16 g^2 s^3 \left(4 g^4-2 g^2 s-3 s^2\right)}{\left(2 g^2-s\right)^2 \left(g^2+s\right)^3 \left(2 g^2+s\right)^2}
\end{multline}
    
\begin{multline}
    \mathcal{G}^{II} : R = \frac{8 s^3 \left(32 g^6-12 g^2 s^2-s^3\right)}{\left(g^2+s\right) \left(2 g^2+s\right)^2 \left(8 g^4+4 g^2 s-s^2\right)^2}
\end{multline}

\begin{multline}
    \mathcal{G}^{III} : R = -\frac{16 s^3 \left(20 g^8-2 g^6 s-84 g^4 s^2-63 g^2 s^3-13 s^4\right)}{3 \left(2 g^2+s\right)^2 \left(18 g^6+5 g^4 s-2 g^2 s^2+2 s^3\right)^2}
\end{multline}

\begin{multline}
    \mathcal{G}^{II}_{mod} : R = \frac{16 s^4 \left(4 g^2+s\right)}{\left(2 g^2-s\right) \left(2 g^2+s\right) \left(8 g^4+4 g^2 s-s^2\right)^2}
\end{multline}

\begin{multline}
    \mathcal{G}^{III}_{mod} : R = \frac{16 s^3 \left(128 g^8+104 g^6 s+24 g^4 s^2-10 g^2 s^3-7 s^4\right)}{3 \left(2 g^2-s\right)^2 \left(2 g^2+s\right)^2 \left(8 g^4+4 g^2 s-s^2\right)^2}
\end{multline}
% }

\bibliographystyle{spphys}       % APS-like style for physics
\bibliography{template-epjc}   % name your BibTeX data base

% \begin{thebibliography}{}
% %
% % and use \bibitem to create references. Consult the Instructions
% % for authors for reference list style.
% % Format for Journal Reference
% Author, Article title, Journal, Volume, page numbers (year)
% Author, Book title, page numbers. Publisher, place (year)
% \end{thebibliography}

\end{document}